\documentclass[12pt,preprint]{aastex}

\begin{document}

\title{Improved Photometric Redshifts with Surface Luminosity Priors}

\author{Lifang Xia\altaffilmark{1}, Seth Cohen\altaffilmark{1},
Sangeeta Malhotra\altaffilmark{1}, James Rhoads\altaffilmark{1},
Norman Grogin\altaffilmark{1}, Nimish P. Hathi\altaffilmark{2},
Rogier A. Windhorst\altaffilmark{1},
Nor Pirzkal\altaffilmark{3}, Chun Xu\altaffilmark{3}}
\altaffiltext{1}{School of Earth and Space Exploration, Arizona State University, AZ, 85287-1404;
lifang.xia@asu.edu}
\altaffiltext{2}{Department of Physics and Astronomy, University of California, Riverside, CA, 92521}
\altaffiltext{3}{Space Telescope Science Institute, Baltimore, MD, 21218}

\begin{abstract}

We apply Bayesian statistics with prior probabilities of galaxy
surface luminosity (SL) to improve photometric redshifts. We apply the
method to a sample of 1266 galaxies with spectroscopic redshifts 
in the GOODS North and South fields
at $0.1\lesssim z\lesssim2.0$. We start with spectrophotometric redshifts (SPZs)
based on Probing Evolution and Reionization Spectroscopically grism spectra, 
which cover a wavelength range of
6000--9000{\AA}, combined with $(U)BViz(JHK)$ broadband photometry
in the GOODS fields. The accuracy of SPZ redshifts is estimated to
be $\sigma (\Delta(z))=0.035$ with an systematic offset of --0.026,
where $\Delta(z)=\Delta z / (1+z)$, for galaxies in redshift range
of $0.5\lesssim z\lesssim1.25$. The addition of the SL prior probability 
helps break the degeneracy of SPZ
redshifts between low redshift 4000 {\AA} break galaxies and high-redshift 
Lyman break galaxies which are mostly catastrophic
outliers. For the 1138 galaxies at $z\lesssim1.6$, the fraction of galaxies
with redshift deviation ${\Delta (z)}>0.2$ is reduced from 15.0\% to
10.4\%, while the rms scatter of the fractional redshift error does
not change much.

\end{abstract}

\keywords{Method: Bayesian Statistic - Prior Probability - Galaxies:
Photometric Redshift - Surface Luminosity}

\section{Introduction}

In recent years, the technique of photometric redshift has been widely used to determine redshifts of 
galaxies for large imaging sky surveys \citep[e.g.,][]{wolf03,mobasher04,mobasher07}. This technique is  
useful for redshift estimation of large numbers of faint galaxies at high redshift which are currently 
too faint for spectroscopy. There are typically two methods of redshift estimation by broadband 
photometry. One approach is an empirical method, which calibrates an empirical training relation
between photometric magnitudes or colors and galaxy spectroscopic redshifts, and applies it to the 
observed photometric sample \citep{connolly95,wang99}. 
Another approach is a template spectral energy distribution (SED) fitting method, which obtains 
best-fit redshifts by comparing the observed SEDs to that of a large empirical or model template 
library \citep{baum62,koo85,fernandez99,bolzonella00,budavari99,budavari00,budavari01,csabai00,
wolf01,blanton03}. The efficiency of SED 
fitting is based on fitting the overall shape of spectra, the detection of strong spectral 
properties, such as the 4000{\AA}/Balmer break and Lyman break, and the amount of dust present in
red galaxies.

The general accuracy of photometric redshift ranges from $\sigma_z=0.02$ to 0.05, which strongly depends on
the number of filters and other factors, such as the precision of the photometry, the zero points, 
the image FWHM, and of course the quality of the templates and the fitting code. 
\citet{hickson94} show that the redshift accuracy by SED fitting is comparable
to slitless spectroscopy from a simulation of 40 band photometry. Practical multicolor sky surveys,
such as the Classifying Objects by Medium-Band Observations (COMBO)-17  survey, using 17
intermediate-band filters \citep{wolf03} and the Beijing-Arizona-Taipei-Connecticut (BATC) sky survey, using 15
intermediate-band filters, achieve a typical accuracy of $\sigma_z=0.02$ for photometric redshift estimation
\citep{zhou01,xia02}. The photometric redshift accuracy using by five broadband filters is about
0.05 \citep{blanton03}. However, the depth of intermediate-band sky surveys are generally constrained to
$z<0.1$, and the observations of  multiple bands can be quite time consuming. Broadband photometry
has the advantage of sensitivity which enables photometric redshifts of large samples of faint and high redshift galaxies.
The photometric redshifts from broandband fluxes tend to have large dispersion and strong degeneracy between low-redshift 
Balmer break galaxies and high-redshift Lyman break galaxies, which leads to the degeneracy of the photometric redshift 
estimation.

To break such degeneracies, \citet{benitez00} developed a Bayesian method of photometric redshift estimation (BPZ) 
using galaxy magnitudes as Bayesian priors. This method produced an accuracy of $\sigma(\Delta(z))\approx0.06$, where 
$\Delta(z)={\Delta z\over 1+z}$, for galaxies in Hubble Deep Field North (HDF-N) up to $z<6$. \citet{mobasher07}  
estimate redshifts for galaxies at $z<1.2$ with 16 bands photometry from 3500 to 23000 {\AA} by 
different photometric redshift codes with and without luminosity function (LF) priors. The results give an accuracy of 
$\sigma(\Delta(z))\approx0.031$ and find slight improvement in the redshift estimation with LF priors.

Observed galaxy surface brightness is a promising observational parameter to break the
redshift degeneracy \citep{koo99}. \citet{tolman30} first showed that the surface brightness dims with redshift as
$(1+z)^{-4}$ in an expanding universe independent of the cosmology.  With this sensitive a dependence on (1+$z$), 
 surface brightness should make a good  prior for the redshift estimation. The only caveat is the evolution of 
intrinsic galaxy luminosity per area with redshift. Passive evolution of stellar populations leads to a significant 
brightening of intrinsic luminosities per unit area at higher redshifts (Pahre et al. 1996; Sandage \& Lubin 2001) 
and therefore to a less steep surface brightness redshift relation.

Using surface brightness priors, \citet{kurtz07} provide a redshift estimator by taking the median redshift in 
small bins in galaxy surface brightness-color space. The estimator is applied to the 10\%-20\% reddest galaxies from 
the Smithsonian Hectospec Lensing survey (SHELS), and achieves an accuracy of $\sigma (\Delta (z))=0.025$ 
for $z<0.8$. \citet{wray07} use the five-band Sloan Digital Sky Survey (SDSS) photometry, surface brightness and the S\'{e}rsic index to provide 
improved photometric redshifts in SDSS. They apply seven-dimensional probability arrays for 
spectroscopically confirmed galaxies at $z<0.25$, which yields $\sigma (\Delta(z))=0.025$ for red galaxies and 0.03 
for blue galaxies. \citet{stabenau08} apply surface brightness priors to ground based VIMOS VLT Deep Survey (VVDS) 
and the space-based GOODS \citep{giavalisco04} field from Hubble Space Telescope (HST), and improve 
the bias and scatter by a factor of 2 for galaxies in the range $0.4<z<1.3$ to get a scatter of
 $\sigma(\Delta(z))\approx0.08$. In this paper, we use spectrophotometric redshifts (SPZs) which use low-resolution grism data and
broadband data in the GOODS fields as our starting point \citep{ryan07,cohen09}. The SPZs  have a scatter in 
 $\sigma(\Delta(z))\approx0.03$. We then use color and surface brightness priors, which we adopt the unit 
of luminosity per area \citep{hathi08}, $L_{\sun}/kpc^{2}$, and hereafter we call it surface luminosity (SL) priors, 
to break the redshift degeneracy to derive photometric redshifts for a sample of 1266 galaxies in the GOODS North (GOODS-N) 
and South (GOODS-S) fields with spectroscopic redshifts between  $0.1\lesssim z \lesssim 2.0$.

This paper is organized as follows. We briefly describe the observations, the data, and
the result of the spectrophotometric redshift estimation in Section 2 . The application of color
and SL priors is given in Section 3. The results of redshift estimation with
hybrid of SPZ and surface luminosity priors are illustrated in Section 4. Finally, we 
discuss our results and present our conclusions in Section 5. Throughout this paper, we assume a $\Lambda$CDM
cosmological model with matter density $\Omega_m=0.28$, vacuum density $\Omega_{\Lambda}=0.72$,
and Hubble constant $H_0=100h$ km s$^{-1}$ Mpc$^{-1}$, with $h=0.7$
for the calculation of distances \citep{komatsu09}.

\section{Observation and Data}

We select a sample of 1266 galaxies in GOODS-N and GOODS-S fields which have both spectroscopic
\citep{wirth04,grazian06,vanzella08} and spectrophotometric redshifts \citep{cohen09} to test the application
of SL priors.  Only spectroscopic redshifts with quality
flag $Q = 0$, or 1 (0: very good quality, 1: good quality) are used. These galaxies have both grism spectra,
from the $HST$/Advanced Camera for Survey \citep[ACS][]{ford03} Probing Evolution and Reionization Spectroscopically (PEARS; S. Malhotra, PI)
survey, and optical broadband $BViz$ photometry from $HST$/ACS 
GOODS version 2.0 images \citep{giavalisco04}. The ACS grism spectra cover a wavelength range from 6000
to 9000 {\AA} \citep{pirzkal04} for objects in parts of the GOODS-N and GOODS-S fields. The galaxies in our
PEARS sample are located in four ACS pointings in the GOODS-N and GOODS-S fields.
The photometry in the GOODS-N field is supplemented with ground-based $U$-band data from \citet{capak04},
and photometry in the GOODS-S field is supplemented with the $JHK$-band data from VLT ESO/GOODS project
\citep{retzlaff09}. The photometry and the aperture correction between the broad-band data are described
in detail by \citet{ryan07} and \citet{cohen09}. 
Figure 1 shows the histogram of the distribution of galaxy spectroscopic redshifts. 
The redshifts of most galaxies are less than z$\sim$2.0. 
The final sample of 1266 galaxies are selected with spectroscopic redshifts in the range of $0.1\lesssim z\lesssim 2.0$.

\begin{figure}[htbp]
\begin{center}
\figurenum{1} \epsscale{0.9}
\hspace{-1.0cm}
{\plotone{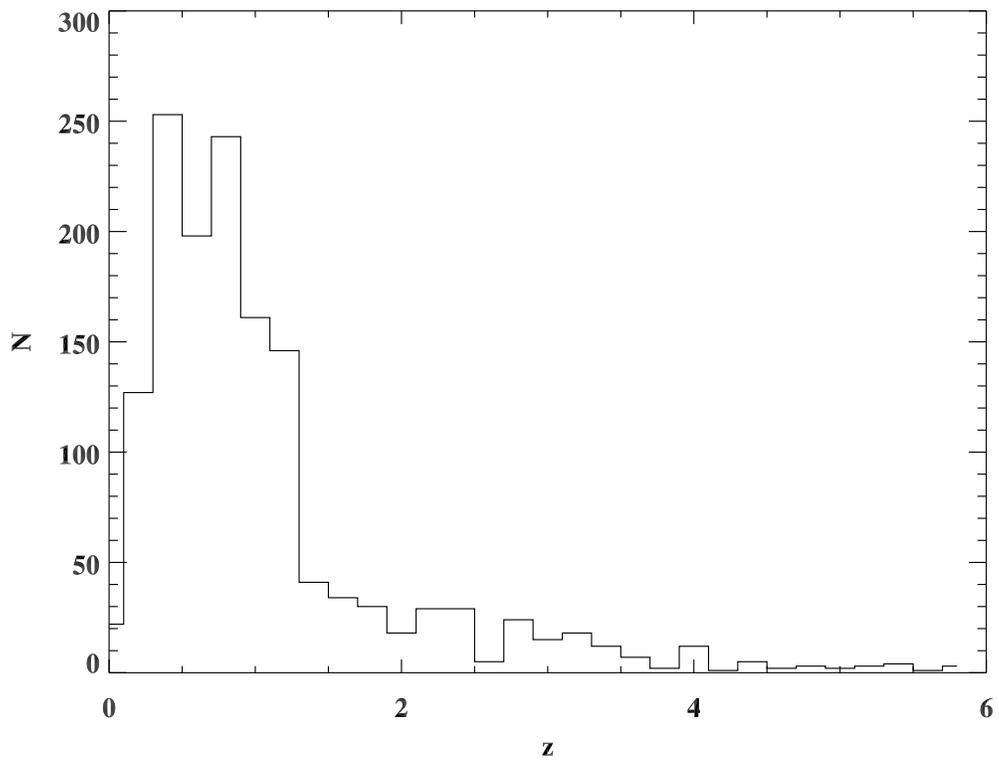}}
\caption{Histogram distribution of the spectroscopic redshifts of the total sample.}
\end{center}
\end{figure}

The SPZs \citep{ryan07,cohen09} are estimated based on the SED fitting of the combination of grism spectra
and UV--optical--infrared broadband photometry by the photometric redshift code $HyperZ$ \citep{bolzonella00}.
The SPZ method achieves a redshift accuracy of $\sigma(\Delta(z))=0.035$ for the 465 galaxies in the
GOODS-N field at redshift range $0.5\lesssim z\lesssim 1.25$ with a catastrophic outlier fraction of 18.2\%. The
catastrophic outliers are defined as galaxies with fractional redshift errors, $\Delta(z)$, greater than 3$\sigma$ of the rms
scatter in the sample. The best accuracy of the SPZ method is achieved for
the redshift range $0.5 \lesssim z \lesssim 1.25$, where the 4000\AA\ break falls in
the peak sensitivity wavelength range of the ACS grism.
The redshifts estimated by SPZ tend to show a strong redshift degeneracy. This is demonstrated in the upper panels
of Figure 6, which compare SPZ redshifts with spectroscopic redshifts. 
A substantial fraction of galaxies at $z<0.6$ scatter to SPZ $\simeq$2 - 3. To improve the redshift accuracy
of the SPZ redshift estimation, we apply the prior probability of galaxy SL to constrain
and break the degeneracy, since surface brightness is tightly related to redshift as approximately
$(1+z)^{-4}$ for bolometric fluxes and $(1+z)^{-(4+\alpha)}$ for fluxes per unit frequency \citep{tolman30}.

\section{Surface Luminosity Priors}

If we were to observe a galaxy with a standard intrinsic luminosity per unit area (hereafter
denoted at $I$) at different redshifts, its measured surface brightness would go down at $I \propto (1+z)^{-4}$.
Due to the limitation of the available photometry in wavelength less than 10,000 {\AA}, 
we choose the rest-fram SL in $B$ band, $I_B$, as prior probability, with
redshifts extending to z$\sim$2.0. The adoption of rest-frame $B$ band is more sensitive to galaxy types from 
starbursts, spirals to ellipticals than redder bands. The intrinsic evolution of galaxy type with redshift
and the observation selection effect will make the relation deviate from power -4 and we 
will calibrate this relation first. A subsample of 283 elliptical galaxies \citep{ferreras09} is used to examine the
difference of the relation between the SL and redshift galaxy types.
For galaxies with redshifts $z\lesssim 0.33$ we 
measure the SL $I_B$ in the band closest to $B$: the $V$ band magnitude for
galaxies at redshift $0.33\lesssim z\lesssim0.96$, the $i$ band for $0.96\lesssim z\lesssim 1.35$, and
the $z$ band for $1.35\lesssim z\lesssim 2.0$.

The photometry of GOODS version 2.0 catalog is measured in AB magnitudes \citep{oke83}, which are defined as:
\begin{equation}
m=-2.5 \log{f_{\nu}}-48.6,
\end{equation}
where $f_{\nu}$ is the flux per unit frequency in unit of erg$\;$s$^{-1}\;$cm$^{-2}\;$Hz$^{-1}$. The half-light
radii are measured by SExtractor and translated into angular radius,  $r_{e}$ (in arcsecond), by
multiplying with the pixel scale $0.\hspace{-0.1cm}^{\prime\prime}03$ pix$^{-1}$.
With the flux $f_{\nu}$ and half-light radius $r_{e}$ in the corresponding band for different redshift range galaxies,
the rest-frame $B$-band SL is calculated as follows
\begin{equation}
I_B=\frac{\Delta\nu_{B} f_{\nu}4\pi d_{L}^2}{(1+z)2\pi d_{A}^2r_{e}^2}=\frac{2\Delta\nu_B f_{\nu}(1+z)^3}{r_{e}^2},
\end{equation}

where $z$ is the redshift of galaxy, $\Delta\nu_{B}$ is the frequency interval corresponding to 
the wavelength range in the $B$ band, $f_{\nu}$ is the flux in the observed filter band, 
$d_L$ is luminosity distance and $d_A$ is angular distance of galaxy,
and $I_B$ is SL in luminosity per unit area (in $L_{\sun}/kpc^2$).
Figure 2 shows the distribution of the rest-frame $B$-band surface brightness with redshift for the 
spectroscopic galaxies. The range of $I_{B}$ goes approximately from $10^{6}$ to $10^{10}L_{\sun}/kpc^2$. 
The upper and lower limits of the observed surface brightness in magnitude per square arcsecond, 
22.3 mag/arcsec$^2$ and 26.3 mag/arcsec$^2$ (corresponding to the magnitude range from 21 
to 25 mag), are plotted as dotted lines in the figure. The relation between log$I_B$ and log$(1+z)$ 
is fitted linearly, which goes as $\log{I_B}=2.61(\pm0.06)\cdot \log(1+z)+6.64(\pm0.01)$. 
The triangular points in the figure represents the elliptical galaxies in our sample. The redshifts
of ellipticals range from 0.3 to 1.4. The ellipticals show generally higher surface luminosities than
blue galaxies while a much similar slope of 2.90($\pm0.6$). Compared with that found in 
\citet{stabenau08}, for passively evolving red galaxies, the observed surface brightness is close to
$(1+z)^{-4}$, and the blue galaxies have a shallower slope, we do not find relatively flatter slope of the 
rest-frame SL for early-type galaxies here, and it may be due to the relatively small 
number of the sample.  The final results show that there is little difference of the improvement in redshift 
estimation accuracy for red galaxies and blue galaxies.

\begin{figure}[htbp]
\begin{center}
\figurenum{2} \epsscale{0.9}
\hspace{-1.0cm}
{\plotone{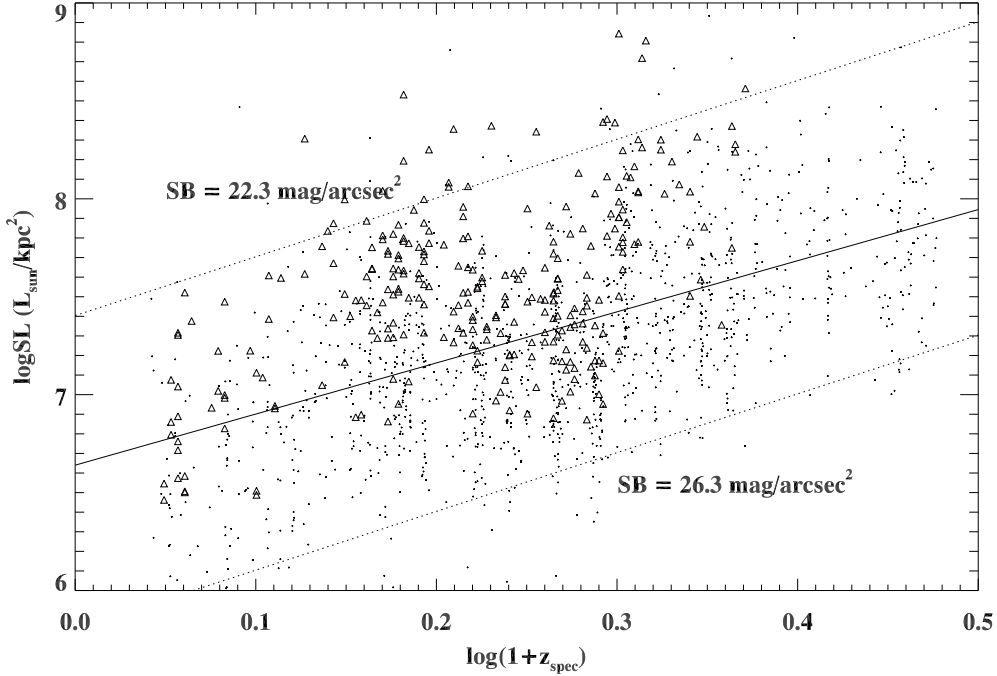}}
\caption{Distribution of rest-frame $B$-band SL $I_B$ as a function of redshift for the total sample. 
The triangular points represent elliptical galaxies in our sample. The upper and lower limits of 
the observed surface brightness, 22.3 and 26.3 mag/arcsec$^{-2}$, are plotted as dotted lines.
The points shows a good linear relation, $\log{I_B}\sim2.61\cdot \log(1+z)$, between SL and redshift.
The ellipticals have a similar slope of 2.90.}
\end{center}
\end{figure}

To apply the scaling of SL with redshift as prior probability, we use
a color-shape \citep{koo85} parameter $(B-V)-(i-z)$ to divide the sample into subsamples.
Figure 3 plots the distribution of $(B-V)-(i-z)$ with redshift. We can see that this shape
parameter declines linearly with redshift at $z\lesssim1.3$ and it increases linearly with
redshift at $z\gtrsim1.3$. This is because that the shape parameter traces the position of 4000
{\AA} break. Three subsamples are obtained with $(B-V)-(i-z)>0.65$, $0<(B-V)-(i-z)<0.65$,
and $(B-V)-(i-z)<0$, corresponding to galaxies in redshift bins of $z\lesssim1.0$, $0.6\lesssim z\lesssim1.2$, and $z\gtrsim1.0$.
The surface luminosity distribution is fitted by Gaussian functions for the three subsamples.
The distribution of log$I_B$ with redshift and the Gaussian fits are plotted in Figure 4.
The peak value of the Gaussian distribution slightly increases from $\log{I_{B,0}}=$ 7.02, 7.03 
to 7.30 with 1 $\sigma$ width of 0.48, 0.49, and 0.44 for the three subsamples, respectively.

\begin{figure}[htbp]
\begin{center}
\figurenum{3} \epsscale{0.9}
\hspace{-1.0cm}
{\plotone{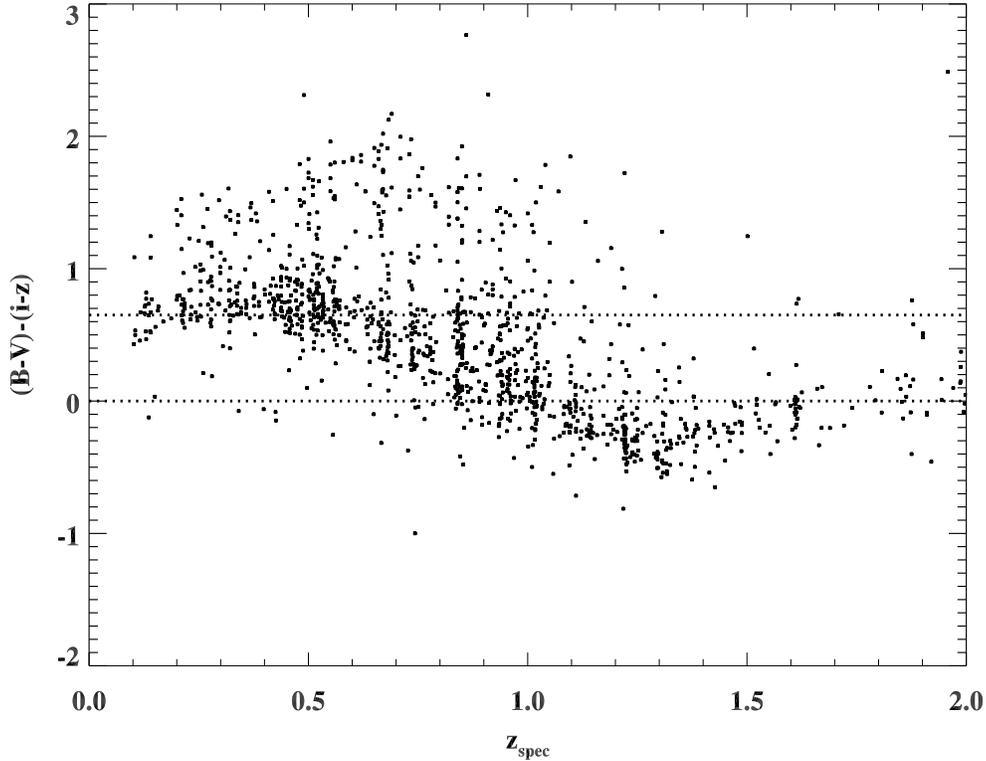}}
\caption{Distribution of color-shape parameter $(B-V)-(i-z)$ with redshift $z$ for the spectroscopic galaxies.
The dotted lines represent the criteria of $(B-V)-(i-z)>0.65$, $0<(B-V)-(i-z)<0.65$ and $(B-V)-(i-z)<0$, which are
implemented to divide sample into three redshift bin subsamples.}
\end{center}
\end{figure}

\begin{figure}[htbp]
\begin{center}
\figurenum{4} \epsscale{0.9}
\hspace{-1.0cm}
{\plotone{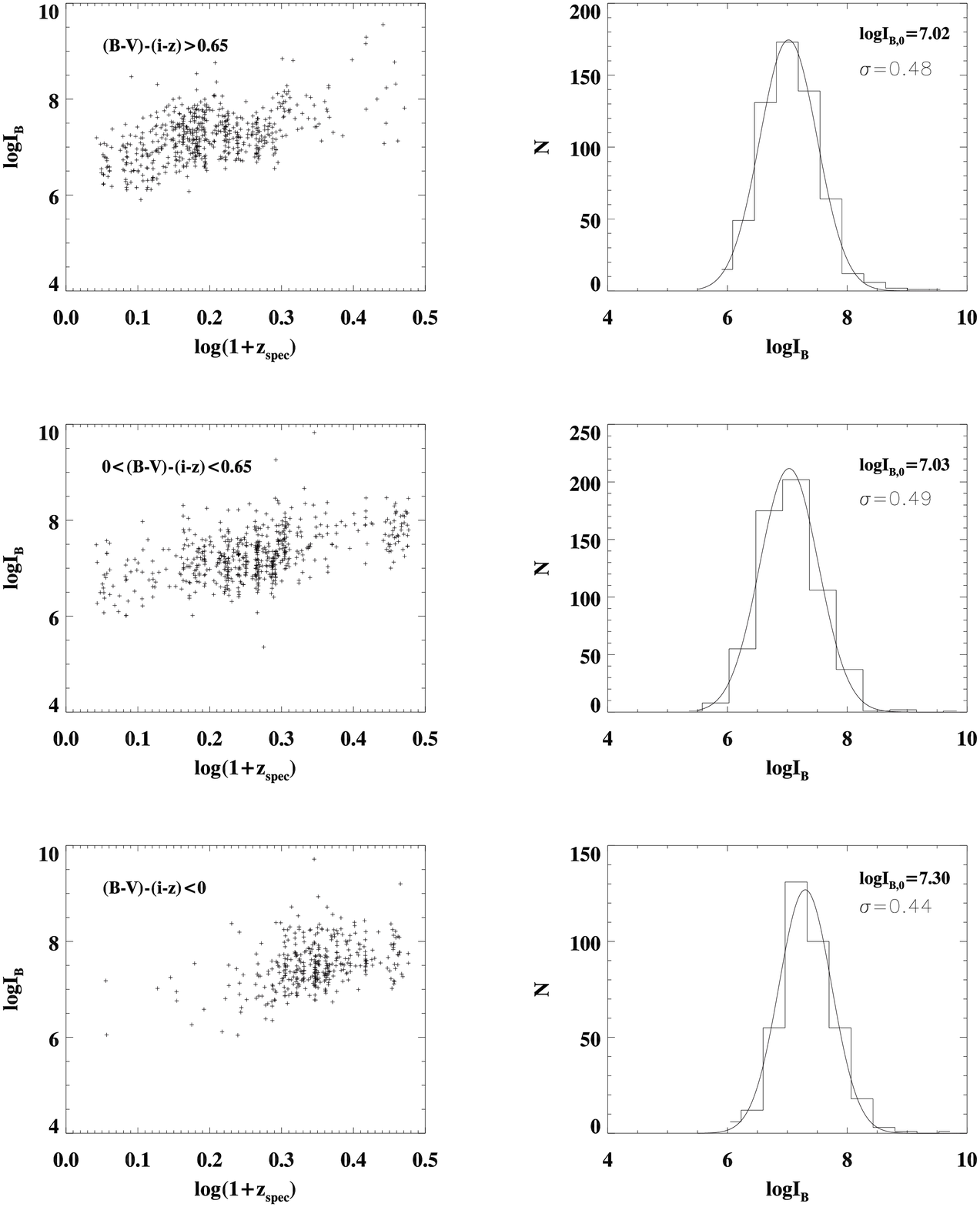}}
\caption{Distribution of log$I_B$ with redshift and the Gaussian fitting for the three color-shape parameter
divided subsamples. Left panel: from top to bottom, the three subsamples have $(B-V)-(i-z)<0.65$,
$0<(B-V)-(i-z)<0.65$ and $(B-V)-(i-z)<0$, respectively. Right panel: the distribution of $\log{I_B}$ is fitted
by a Gaussian function. The peak and the width of the Gaussian distributions are $\log{I_{B,0}}=7.02$, 7.03, 7.30
and $\sigma=0.48$, 0.49, 0.44 for the three subsamples, respectively.}
\end{center}
\end{figure}

The SL prior probability is calculated with the formula as
\begin{equation}
p(z|I_B(z))=\frac{\phi}{\sqrt{2\pi}\sigma}\exp(\frac{-(\log{I_B}(z)-\log{I_{B,0}})^2}{2\sigma^2}),
\end{equation}
where $\phi$ is the normalization constant so that the integration of the probability in the studied
redshift range ($0<z<7$) is 7, $\sigma$ is the width of the Gaussian profile,
and $\log{I_{B,0}}$ is the Gaussian peak value. $I_B(z)$ is the surface
brightness for one galaxy
at different redshifts, calculated over a redshift range $0.10<z<7.0$ with a step of 0.005, the same as 
that of SPZs. The best redshift is
estimated by the combination of SL prior probabilities and SPZ fitting probabilities, which are
output from $HyperZ$.  Using Bayes' theorem, the final probability of redshift can be computed as
\begin{equation}
p(z|I_B(z),C)= {{p(z|I_B(z))\times p(C|z)} \over {p(C)}},
\end{equation}
where $p(z|I_B(z))$ is the redshift probability given by SL priors, and $p(C|z)$, 
$P(C|z)=\exp(-\chi^2(z)/2)$, is the
probability of the galaxy at redshift $z$ with the observed color $C$ given by the SPZs estimation. 

\section{Implication and Results}

For the 1266 galaxies, we first divide galaxies into subsamples by the color-shape parameter. Then we calculate
the SL prior probability for galaxies by the corresponding Gaussian profiles in different subsamples. Combining
the SL prior probability with the SPZ likelihood function, we obtain the best redshift as the maximum of
the final probability distribution.

\begin{figure}[htbp]
\begin{center}
\figurenum{5} 
\epsscale{0.9}
{\plotone{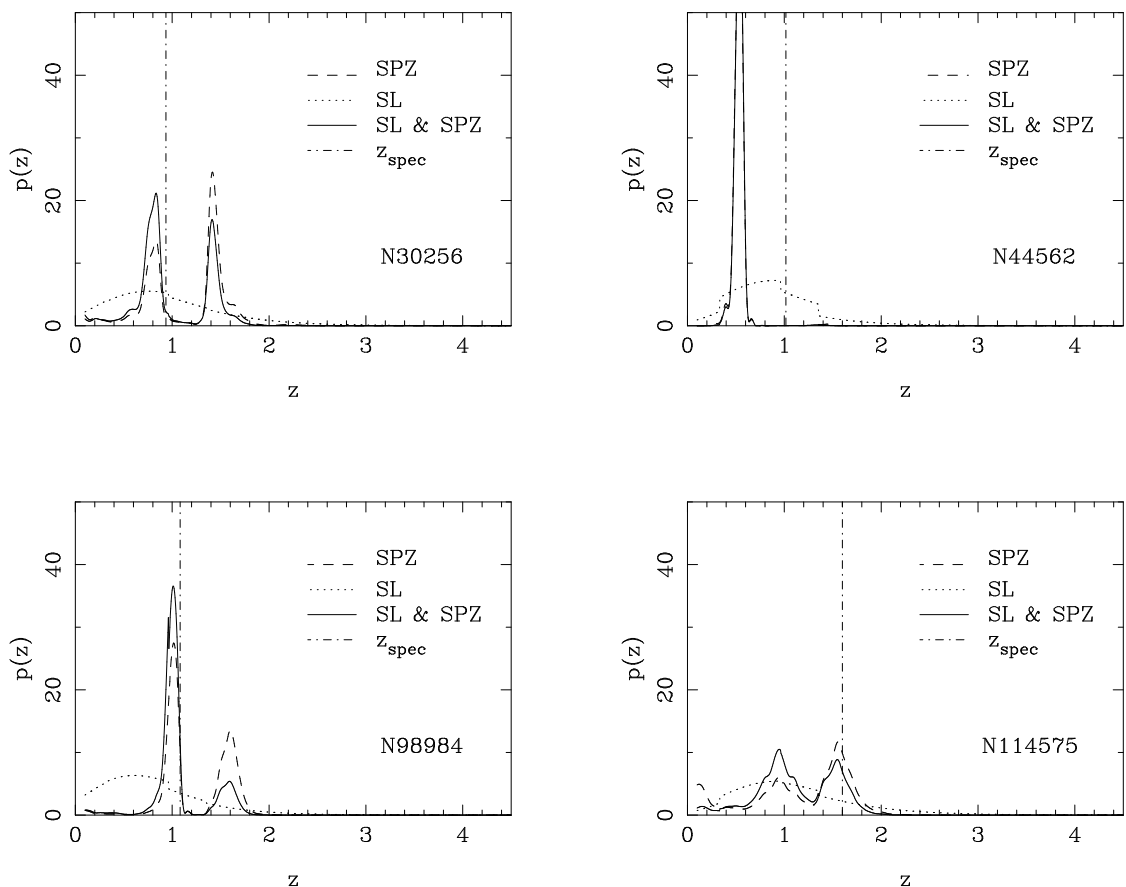}}
\caption{Probability distributions as a function of redshift obtained from the SPZ SED fitting, the SL priors,
and the combination of SPZ SED fitting with SL priors. The dashed line represents the likelihood function
given by SPZ SED fitting. The dotted line represents the calculated probabilities by SL priors.
The solid line shows the combined probability distribution. The vertical dash-dotted line represents the position of 
the spectroscopic redshift. The object ID is labeled at the right-bottom in the panel.}
\end{center}
\end{figure}

Figure 5 shows four examples of redshift probability distributions for galaxies in GOODS-N
field. The ID of the object is labeled at the right-bottom of the panel. The dashed line in the
figure represents the likelihood function given by SPZ SED fitting. The dotted line
represents the calculated probabilities by SL priors. The solid line shows the combined probability distribution
from SPZ SED fitting and SL priors. The vertical dash-dotted line represents the position of the spectroscopic redshift.
The upper-left panel shows a case where the SPZ redshift estimation gives two peaks in the redshift probability function.
The addition of the SL priors probability gives the correct distribution around the spectroscopic redshift. 
With the combination of the two probabilities, the correct peak is chosen, and the probability of a catastrophic 
redshift estimation is reduced. The upper-right panel shows an example where the SPZ does not produce a reasonable 
likelihood distribution, though the SL priors give more reasonable estimation. The lower-left panel gives an example of
the correct estimation of redshift by both methods. In the lower-right panel, the SL priors choose the wrong peak 
of the SPZ $p(z)$ distribution for a galaxy with redshift $z=1.6$. This can be the reason 
of the larger deviation of redshift estimation with SL priors at redshift $z>1.6$. 
The results of redshift estimation with SL priors are shown in Figure 6.

\begin{figure}[htbp]
\begin{center}
\figurenum{6} \epsscale{0.9}
\hspace{-2.0cm}
{\plotone{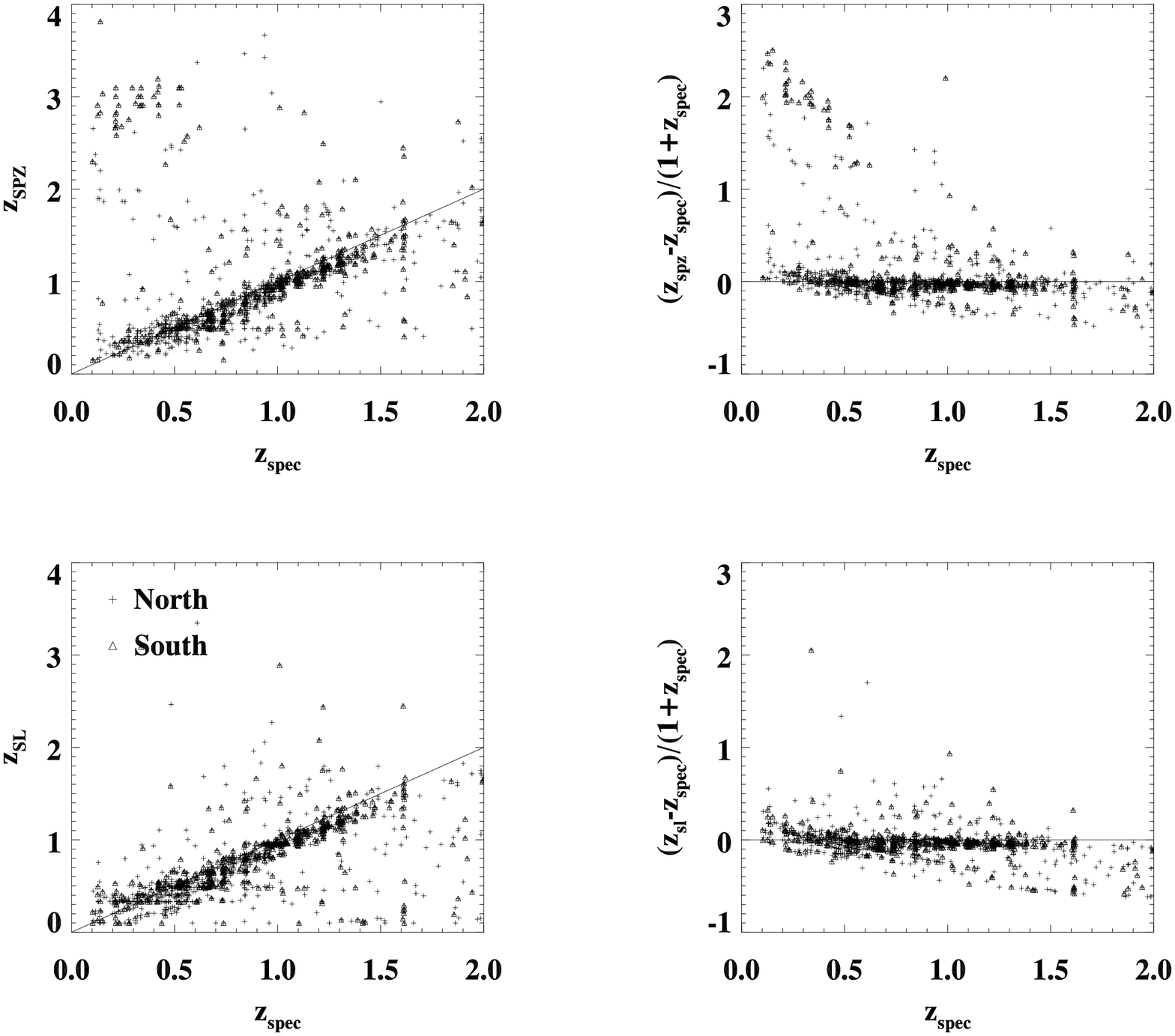}}
\caption{Left panels show the comparison between estimated redshifts and spectroscopic redshifts. The upper
one is the comparison between SPZ redshifts and spectroscopic redshifts. The lower one is that of the improved SPZ
redshifts by SL priors. The cross points illustrate the galaxies in GOODS-N field; the triangular points are galaxies
in GOODS-S field. The right panels show the distribution of the fractional error $\Delta(z)$ with redshift for the redshift
estimation with and without SL priors.}
\end{center}
\end{figure}

Figure 6 shows the comparison of SPZs with and without SL priors.
The upper two panels in Figure 6 show the comparison between SPZ redshifts and spectroscopic redshifts, and the
distribution of redshift fractional error $\Delta(z)={\Delta z\over 1+z}$ with spectroscopic redshifts.
The triangular points in figure are galaxies in the GOODS-S field which are supplemented
with infrared $JHK$ photometry, and the cross dots are galaxies in the GOODS-N field, which have $U$-band data.
From this comparison, we can see that many galaxies in the GOODS-S field with $z<0.6$ are estimated to be around
$z\simeq$ 2--3 by the SPZ method. Because the 4000 {\AA} break of $z<0.6$ galaxies falls in the $UV/B$ band, 
it can be confused with galaxies of $z \sim 3.0$ with the Lyman break falling in $B/V$ band. From this comparison,
we can also see that the scatter improves greatly for galaxies in GOODS-N field. The GOODS-N field has fewer
catastrophic outliers because of $U$-band photometry for galaxies.

The bottom two panels show the results of the photometric redshifts with SL prior probabilities.
From the comparison of the redshift estimation with and without SL priors, the effectiveness of SL priors
is illustrated in breaking the redshift degeneracy, and in reducing the fraction of catastrophic
outliers. For the total sample at $0.1\lesssim z\lesssim2.0$, the accuracy of the redshift estimation by SL priors
(which is the width of the Gaussian error distribution) changes little from 
$\sigma(\Delta(z))=0.043$ with an systematic offset of --0.019 to
$\sigma(\Delta(z))=0.044$ with an offset of --0.020. We can see from the figure that at redshifts
$z\lesssim0.3$ and $z\gtrsim1.6$, the SL priors do not work as well as in the intermediate redshift range. This is because the
peak value of the SL sampled by the SL priors is slightly larger than the actual SL for
galaxies with lower redshifts, and is slightly smaller than the actual SL for the galaxies with highest
redshifts. For galaxies in the redshift range $0.5\lesssim z\lesssim1.25$, the rms error remains the same at
$\sigma_x=0.035$ for both methods.  For galaxies with redshift
$z>1.6$, the SPZ yields large scatter. We only use the 1183 galaxies at $z<1.6$ to calculate the statistics of 
catastrophic outliers.  For galaxies with $|\Delta(z)| >0.2$,
the fraction decreases from 15.0\% to 10.4\% by adding SL priors; and for galaxies with 
$|\Delta(z)| >0.5$, the number reduces from 87 to 22. This effect is demonstrated clearly in Figure 7, which shows 
the histogram of the fractional redshift error. The solid line shows the histogram of galaxies with
improved SPZ redshifts by SL priors. The dotted line represents that of galaxies with SPZ redshifts.
We can see that the galaxies with fractional errors greater than 0.6 almost disappear with the SL
priors method.

For the 283 elliptical galaxies, the redshift estimation shows same trend as that of the total sample,
with little change in accuracy and improvement in catastrophic outliers. The redshift accuracy is $\sigma(\Delta(z))\sim$0.01,
much better than that of the blue galaxies, for both SPZs and SPZs with SL priors. The elliptical
galaxies in the spectroscopic sample is not complete due to the selection effects and it can lead to small
difference in the accuracy estimation. In the application to the photometric sample with this calibration, 
there is type selection bias in different redshifts. At higher redshifts, the photometric sample tends to 
have more luminous elliptical galaxies, which should have better accuracy in redshift estimation.

\begin{figure}[htbp]
\begin{center}
\figurenum{7} \epsscale{0.9}
\hspace{-1.0cm}
{\plotone{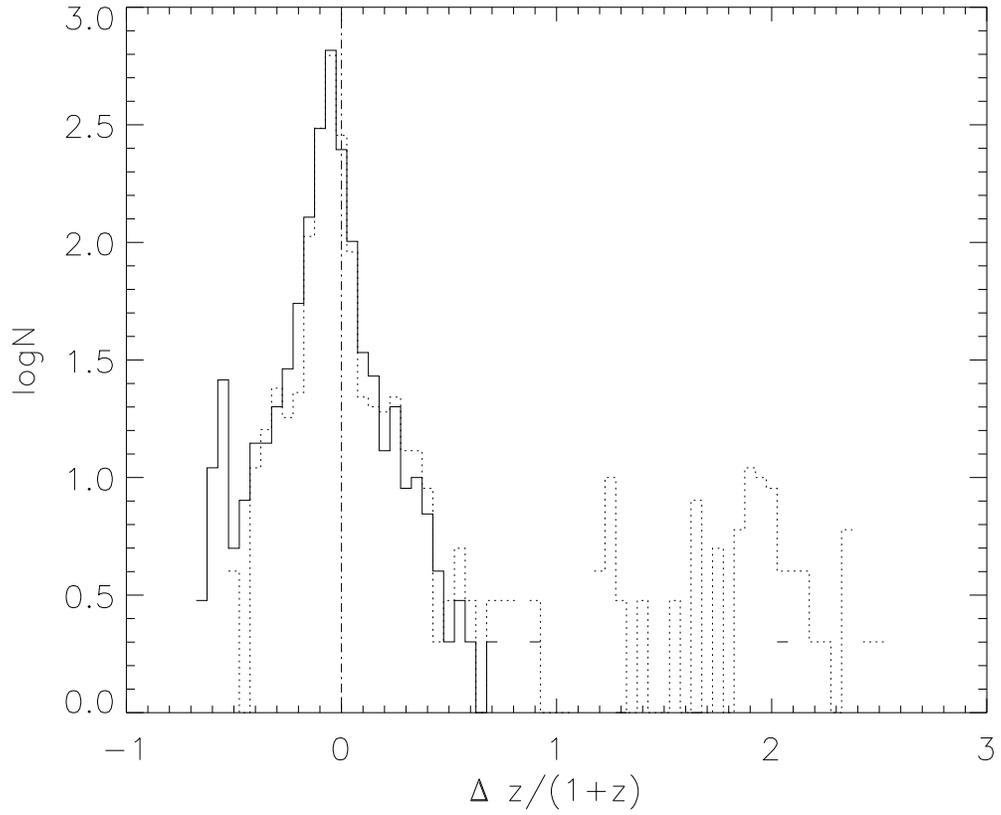}}
\caption{Histogram distribution of redshift fractional error $x$. The solid line shows
the improved SPZ redshifts by SL priors. The dotted line represents the distribution of SPZ redshifts.
The dash-dotted line shows the zero position of the histogram.}
\end{center}
\end{figure}

The color and SL priors works well for lower redshift samples at $z\lesssim1.6$. However, to 
apply this method to redshift estimation for the whole PEARS sample, we need to
improve the method, since the whole sample will include such galaxies at
$z\gtrsim1.6$ and the relation between the shape parameter $(B-V)-(i-z)$ and redshift will not be
near linear. The value of $(B-V)-(i-z)$ will go up linearly with redshift at $z\gtrsim1.6$. The
application of this method needs to be studied further, likely with additional near-IR filters to
obtain the photometry of rest-frame $B$-band.
This can be done with the $HST$/WFC3 after 2009.

\section{Summary and Conclusions}

For an object with constant luminosity per unit area, the bolometric surface brightness scales
as $(1+z)^{-4}$ in an expanding universe. That, combined with the fact that there is a definite
upper limit to luminosity per unit area seen in starburst galaxies from $z=0$--$7$ 
\citep{hathi08, meurer97}, would make for a very strong prior for photometric estimates. 
However, the mean luminosity per unit area is well below this upper limit and shows strong
redshift evolution for blue late-type galaxies. The early-type galaxies show
a generally higher SL and a similar slope of the redshift evolution.

To calibrate the evolution of luminosity per unit area, we divide the sample into three redshift
bins using a color-based criterion; and then derive the distribution of luminosity per unit area
in rest-frame $B$ band. The probability of the rest-frame SL is applied 
as a prior to the redshift probabilities given by SED fitting to broadband +  grism data.

The method is applied to 1266 galaxies observed with $HST$/ACS PEARS grism spectra and with GOODS $BViz$
broadband photometry and known ground-based redshifts in the range of $0.1\lesssim z\lesssim2.0$. The 
accuracy is assessed with the spectroscopic redshifts. By comparing the redshift estimation with and 
without SL priors, the new method improves the number of galaxies with $|\Delta(z)|>0.2$ from 15.0\% 
to 10.4\%. The rms scatter does not change much. The improvement seems same for the blue galaxies and
the 283 red galaxies, while the red galaxies show higher accuracy in redshift estimation. 
The result shows the efficiency of the SL priors in breaking the degeneracy of SPZ redshifts for low-redshift 
Balmer break galaxies and high-redshift Lyman break galaxies.

\begin{acknowledgements}

PEARS is an HST Treasury Program 10530 (S. Malhotra, PI). Support for program was provided
by NASA through a grant from the Space Telescope Science Institute, which is operated by
the Association of Universities for Research in Astronomy, Inc., under NASA contract
NASA5-26555 and is supported by HST grant 10530.

\end{acknowledgements}

\end{document}